\documentclass[12pt,a4paper]{article}

 \usepackage[dvips]{graphics}

 \usepackage{afterpage}
 \usepackage{enumerate}
\begin{document}

\title{\bf Modeling and Analysis of a Spectrum of the Globular Cluster NGC~2419}

\author{\bf Sharina M.E.$^{1}$, Shimansky V.V.$^{2}$, Davoust E.$^3$}
\date{November 10, 2012}%
\maketitle
{$^{1}$ Special Astrophysical Observatory, Russian Academy of Sciences,  N.Arkhyz, KChR, 369167, Russia} \\
{$^{2}$ Kazan Federal State University, Kremlevskaya street 18, Kazan, 420008, Russia} \\
{$^{3}$ IRAP, Universit\'e de Toulouse, CNRS, 14 Avenue Edouard Belin, 31400 Toulouse, France} \\
%\maketitle
\begin{abstract}
{NGC~2419 is the most distant massive globular cluster in the outer Galactic halo.
It is unusual also due to the chemical peculiarities found in its red giant stars 
 in recent years. We study  the stellar population of this unusual object using spectra 
obtained at the 1.93-m telescope of the Haute-Provence Observatory.
At variance with commonly used methods of high-resolution spectroscopy
applicable only to bright stars, we employ spectroscopic information on the
integrated light of the cluster. We carry out population synthesis modeling of 
medium-resolution spectra using synthetic stellar atmosphere models based on 
a theoretical isochrone corresponding accurately to the observed color-magnitude diagram. 
We study the influence of non-Local Thermodynamic Equilibrium for some chemical 
elements on our results. 
The derived age (12.6 Gyr), [Fe/H]=-2.25 dex, helium content Y=0.25, and abundances 
of 12 other chemical elements are in a good qualitative agreement with published
high-resolution spectroscopy estimates for red giant members in the cluster.
On the other hand, the derived element abundance, [$\alpha/Fe]=0.13$ dex (the mean of
 $[O/Fe]$, $[Mg/Fe]$ and $[Ca/Fe]$), differs  from the published one ($[\alpha/Fe] =0.4$ dex)
for selected red giants in the cluster and may be explained by a large dispersion in 
the $\alpha$-element  abundances recently discovered in NGC~2419. We suggest that 
studies of the {\it integrated} light in the cluster using high-resolution spectrographs
in different wavelength regions will help to understand the nature of these chemical anomalies.}
\end{abstract}

\section{INTRODUCTION}
Hubble Space telescope (HST) observations and the development of new powerful high-resolution 
spectrographs at large ground-based telescopes demonstrate that GCs do not consist 
of stars of a single age and metallicity (see the review paper by  Gratton et al. (2012) and 
references therein). In addition to stars having a chemical composition similar to that of the
Galactic halo, GCs contain stars built of the matter that has passed through the full 
$CNO$-cycle and proton capture of light nuclei. Contrary to the field stars, 
they often show Sodium - Oxygen anti-correlations (Gratton et al. (2012)). Anomalies
of chemical and stellar composition are found for many Galactic GCs. 
Explaining the origin and chemical evolution of stars in GCs is very important
for understanding the processes of nucleosynthesis and chemical evolution in the early Universe.

Although deep stellar photometry is available for many Galactic GCs thanks to 
HST high-resolution images, high resolution spectroscopy in a wide spectral range 
for many faint stars in GCs is still very expensive. 
 This is why we decided to explore the possibility of studying the detailed chemical composition in GCs using  medium-resolution spectra. 
In the present paper, we fit medium-resolution spectra of NGC2419 integrated along different 
slit positions with stellar atmosphere models taking into account the stellar mass function.

NGC2419 is a unique massive GC ($M_V=-9.^m5$) in the outer Galactic halo, at a distance of 84 kpc. 
Its large half-light radius ($R = 23$ pc, Ibata et al., 2011) is similar to that of ultra-compact
dwarf galaxies. The stellar content and mass function, 
the chemical composition and the structure of NGC~2419 
have been studied actively. The blue horizontal branch (BHB) of the cluster is very extended 
(di Criscienzo et al. (2011)). 
The stellar mass function and the properties of horizontal branch (HB) stars, Hertzsprung gap stars 
and blue stragglers have been explored in a number of papers using deep color-magnitude diagrams ($CMD$) 
(Harris et al. 1997, Dalessandro et al. 2008, di Criscienzo et al. 2011, Bellazzini et al. 2012). These studies gave information about the
age and metallicity of NGC~2419. The Helium and alpha-element abundances 
were estimated in the aforementioned papers by comparison of the $CMD$ of NGC~2419 
with model isochrones (Pietrinferni et al. (2004, 2006)). 
An unusually high mass-to-luminosity (M/L) and the possible presence 
of a black hole in the center of the cluster were suggested by investigators of its stellar dynamics 
(Ibata 2011a,b). Measurements of stellar velocities at higher spectral resolution (Baumgardt et al., 2009)
did not favor the last hypothesis. So far detailed chemical compositions are
known only for  brightest red giant stars of NGC~2419 (Cohen et al. (2011), Shetrone et al. (2001)).
In the latter papers the unusual chemical composition and other properties of the cluster were interpreted 
as a signature of its possible extragalactic origin.

In this paper we carry out population synthesis modeling of medium-resolution spectra of NGC2419. 
Our main aim was to obtain theoretical spectra in good agreement with the observed ones.
This task is solved for the first time. 
The lowest resolution used in the literature for such a task was $R\sim25000$ 
(Colucci et al., 2012).
At variance with the population systhesis modeling of  Schiavon et al. (2002), 
we use theoretical spectra of stars calculated using stellar atmosphere models. 
Medium resolution allows us to extend the method to extragalactic GCs
at distances of $\sim 10$ Mpc.  

In section 2 we describe our observational material and procedure of spectra reduction.
In section 3 we explain our method of theoretical population synthesis. 
In section 4 we derive the age and metallicity of NGC2419 and a mean Helium abundance by
comparison of theoretical and observed Balmer line profiles.
In section 5 we describe how we derive chemical abundances for the main chemical elements 
by fitting absorption-line features in the observed spectra
with theoretical ones. Our conclusions are formulated in section 6.

\section{OBSERVATIONS AND DATA REDUCTION}

Medium-resolution long slit  ($5.5' \times 2''$) spectra of  NGC~2419 were obtained
with the spectrograph CARELEC (Lemaitre et al. (1990)) at the 1.93-m telescope of 
the Haute-Provence Observatory.
The total exposure time was 2.6 h (Table 1), with a seeing of 
$\sim 2.5'' \div 3'' $. The grism 300 grooves/mm and the slit width $ 2'$ provided a resolution of
$\sim 1.78$\AA\ pix$^{-1}$, FHHM$\sim 5$\AA\ and spectral range $3700 \div 6800$\AA.
He and Ne lamps were exposed at the beginning and the end of each night for wavelength calibration.
The spectrophotometric standard star HR1544 (Hamuy et al., 1992, 1994) was observed for
relative calibration of the spectral sensitivity of the spectrograph.
Lick standard stars (Worthey et al. (1994), Table 1) were observed each night for
calibration of radial velocities and spectrophotometric indices. Note that
our observing program also included other distant Galactic GCs. Separate papers
will be devoted to their spectroscopic analysis.

Data reduction and analysis were carried out using MIDAS
 (European Southern Observatory Munich Image Data Analysis System) (Banse et al., 1983)
and IRAF (Image Reduction and Analysis Facility) {\footnote {http://iraf.noao.edu/}}.
Cosmic-ray rejection was made using the program {\it filter/cosmic} in
MIDAS. A map of hot pixels was prepared prior to the masking process.
The accuracy of the dispersion solution obtained using the He-Ne lamp is  $\sim 0.08$\AA.
Possible systematic uncertainties may arise due to differences in optical paths from the lamp
to the spectrograph and from a position on the sky to the spectrograph.
Such uncertainties were corrected using wavelengths of prominent emission night-sky lines, 
e.g. $[OI] 5577$ \AA. One-dimensional spectra were extracted using the procedure {\it apsum} in IRAF. 
This program allows one to correct the curvature of spectra along the dispersion, which may arise due to a
slight slope of the slit, or as a result of the influence of atmosphere dispersion if
the slit position angle has a large parallactic angle (PA>> 0).

The signal-to-noise ratio (S/N) achieved in the summed one-dimensional spectrum is
$ S/N \sim 130$ in the middle of the spectral range. 
The heliocentric radial velocity of NGC~2419 derived  by cross-correlation using
radial velocity standard stars and the program {\it fxcor} in IRAF agrees well with the literature one:
$V_h=-20$ km s$^{-1}$ (Harris, 1996){\footnote{http://physwww.mcmaster.ca/~harris/mwgc.dat}}.
For calculating this value using bright night-sky emission lines 
we took into account the heliocentric correction (Stumpff, 1980) 
and possible systematics of our wavelength scale with respect to the standard one.

\section{MODELING OF INTEGRATED LIGHT SPECTRA OF GLOBULAR CLUSTERS}

\subsection{{\it Selection and optimization of isochrones}}

Modeling the GC integrated light was based on 
synthetic spectra for stars with physical parameters corresponding to
the selected isochrone. We use isochrones of  Bertelli et al. (2009)
having a wide ranges of age ($t$), heavy element abundance ($Z$) and Helium abundance ($Y$).

After initialization of a specified isochrone, 
its optimization was carried out to reduce the amount of calculations.
For the stars (places of isochrones) 
which contribute less than $0.5\%$ to the cluster light 
we use the following maximum  steps of temperature
 and surface gravity: $\Delta \log(T_{eff}) = 0.01$ and $\Delta \log~g = 0.06$.
For the other points the steps were twice as large. The places of isochrones with the initial 
stellar masses which contribute less than $\Delta M < 10^{-4} M_{\odot}$
to the cluster mass were excluded from the calculations.
Finally the resulting isochrones contained 55-65 points.
So possible errors of modeling caused by the optimization procedure were reduced
to $0.005\%$ of the continuum flux.

\subsection{{\it Theoretical spectra of stars}}
We used a modified $SPECTR$ package 
( Shimansky et al., 2003) for calculating synthetic blanketed spectra of stellar atmospheres.
Plane-parallel, hydrostatic stellar atmosphere models
for a given initial set of parameters  ($T_{eff}$, $\log~g$, $M/H$) 
were obtained by interpolation of a model grid of Castelli \& Kurucz (2004)
using a method described in Suleimanov (1996). 
We used {\it the solar} chemical composition and isotopic abundance of some chemical elements
(Anders \& Grevesse, 1989).

For each stellar atmosphere we derived fluxes propagating in three basic directions
with angles to the stellar surface $\theta=62^o$, $30^o$, $8^o$.
Then the stellar surface was divided into sectors with a step of $5^o$ 
for all orienting angles. For each corner
the area and the angle of visibility were calculated.
The radiation intensity of the sector reaching an observer was found by
interpolation of the specific intensities in the three basic angles according 
to the actual visibility angle. The integrated stellar radiation was obtained by
integration of the radiation from each sector according to the area and local 
radial velocities caused by the rotation of the star and radial
tangential macro-turbulence. Finally, the obtained spectra were 
broadened according to the instrumental spectral-line profile of the spectrograph, 
using a Gaussian function with a half-width corresponding to the resolution.

Before normalization,
spectra with and without atomic lines and molecular bands were calculated simultaneously
in the spectral range $\lambda=3900$--$5900$\AA. 
Modeling the spectra with different wavelength steps  $\Delta\lambda$ has shown that
 $\Delta\lambda\le 0.05$\AA\ is the maximum value when errors in the line
profiles are less than $0.005\%$ of the continuum flux. 
We took into account all sources of continuous opacity tabulated in the packages
$STARDISK$ (Suleimanov, 1992) and $SPECTR$ (Sakhibullin \& Shimansky, 1997), approximately 570000
atomic lines from the list of Kurucz (1994) and 28 molecular bands in the 
optical spectral range  calculated according to the theory of  Nersisyan et al. (1989)
kindly provided by Ya.V. Pavlenko.
We used the empirical oscillator strengths calculated by  Shimansky (2011)
for the 750 most intense lines.

Additionally, we found that it was necessary to take into consideration the $CH$ molecule,  
powerful bands of which are seen in the spectrum of NGC~2419 in the range 
$\lambda =4100$--$4500$\AA. Therefore we included  calculation 
of the opacities for 75000 molecular bands $^{12}CH$ and $^{13}CH$ 
from the catalogs of Kurucz (1994) in the package $SPECTR$.
Modeling of the Balmer-line profiles was done using of the Stark broadening theories
of  Vidal, Cooper \& Smith (1973) and  Griem (1960).
For the other lines we calculated the standard Voigt profile taking into account the 
following information:  Doppler broadening of lines due to thermal motion of atoms, 
microturbulent velocity $\xi_{turb}$,  natural damping, 
 Stark broadening according to the approximation
of  Kurucz \& Furenlid (1979) and the van der Waals broadening constants  defined using 
the classical formula  Unsold (1955)
 with the the scaling factor in the range $\Delta \log C_6 = 0.7$--$1.2$ (Shimansky, 2011).

\subsection{{\it Non-LTE effects}}

We modeled the line profiles  of chemical elements taking into account departures
from Local Thermodynamic Equilibrium (LTE)  according to the methods developed by 
Ivanova et al. (2004) and  Shimansky et al. (2008) for binary stellar systems.  
The non-LTE atomic populations for each atmosphere model were calculated using 
 the linearization method by  Auer \& Heasley (1976)
with the package $NONLTE3$ of Sakhibullin (1983).
We used:
 $23$-level model of the $ H I$ atom (Shimansky et al., 2008), $21$-level model of the $Na I$ atom (Mashonkina et al., 2000),
$50$-level model of the $Mg I$ atom (Shimansky et al., 2000), $45$- level model of the $Mg II$ ion
(Shimansky, 2002), $39$-level model of the $Al I$ atom (Menzhevitski et al., 2012), $36$-level model of the
 $K I$ atom (Ivanova et al., 2000) and $43$-level model of the $Ca II$ ion (Ivanova et al., 2002). 
The influence of  non-LTE effects on line profiles for different chemical elements
in the summed spectrum of NGC~2419 is shown in Fig.1
for the physical parameters and chemical composition obtained by us (see Sec.~4).
Note that non-LTE effects do not affect the HI line intensities
(increase by $ < 0.3\%$ of the central line depth) due to the 
dominance of HI in almost all stellar atmospheres of the cluster. 
Lines of heavy elements are enhanced with very different amplitudes taking into account 
non-LTE effects. The intensity of the resonance doublet  $Na I~\lambda\lambda~5889,~5895$\AA\
grows a lot (up to $15\%$ of the central depth). As a result, the sodium abundance is reduced 
by $0.25$ dex. The non-blended resonance line $Ca II~\lambda~3933$\AA\ is
significantly influenced by non-LTE effects (up to $5\%$ of the central depth).
On the other hand, the lines $Mg I$ and $Mg II$ do not change.
Unfortunately, we have no near-IR spectra containing 
the lines $AlI$, $KI$ having significant non-LTE corrections. 

\subsection{{\it Calculation of integrated spectra}}

The calculated spectra for stars belonging to the selected isochrone 
were summed according to their radii and initial mass function $\psi(m)$. 
In this work the   Chabrier (2005) initial mass function (IMF) was used

\begin{displaymath}
\psi(m) \propto \exp \left(-\frac {(\log(m) - \log(0.2))^{2}} {2 \times (0.55)^2} \right)~~(m \le 1),
\end{displaymath}

\begin{displaymath}
\psi(m) \propto m^{-1.35 \pm 0.3}~~(m \ge 1),
\end{displaymath}

where $m = \frac {M}{M_\odot}$. It gives a good approximation of
the stellar luminosity function (LF) found in the literature for NGC 2419 (Bellazzini et al., 2012).
The Chabrier IMF matches well the  theoretical mass distributions of  Padoan \& Nordlund (2004) 
for lowest mass stars.
Finally, normalization of the integrated spectrum was done by division of the
spectrum with its lines and molecular bands by the corresponding continuum.

\section{EVOLUTIONARY PARAMETERS OF NGC~2419}

The  isochrone used for the cluster is defined by its parameters (age $t$, helium abundance 
 $Y$ and metallicity $Z$)\footnote{The abundance of chemical elements according to their 
mass is defined as follows: $ X + Y + Z = 1$,
where $X$ corresponds to the hydrogen abundance, $Y$ stands for the Helium abundance, and
 $Z$ means the heavier elements abundance. 
The values for the Sun are $ Y=0.273$ and $ Z=0.019$ (Anders \& Grevesse, 1989)}
which may be derived by comparison of the observed and theoretical spectra.
The index $Z$ of heavy-element abundance is related
to the chemical composition in the atmospheres of stars belonging to the cluster ($[M/H]$):
 $z = 0.019 \times 10^{[M/H]}$. It can be obtained with an accuracy of $\sim 0.2$ dex from the
analysis of synthetic spectra (see below). To define the initial $t$ and $Y$ 
we adopt the metallicity from the literature $Z=0.0001$ (Bellazzini et al., 2012).
Note that chemical elements synthesized in deep stellar layers of some red giant and 
 blue horizontal branch (BHB) stars may be delivered to the surface and
change the initial chemical composition (e.g. Smiljanic et al., 2007). However, 
our analysis shows that an upper limit for the contribution of these stars 
to the total flux  in the spectral range $\lambda = 3900$--$5900$\AA\ is only $6\%$. 
We show in Fig.~2 the contribution of different parts of the isochrone to the total spectrum.
If we assume that  chemical anomalies due to the mixing effect do not exceed $0.5$ dex, 
the errors $\delta Z$ are $\sim 10\%$ (Pereira et al., 2011) and the determined abundances of 
chemical elements for the cluster are close to their initial values.

We derive  $Y$ and $Z$ as follows. We vary them until
 the theoretical profiles for the lines $H_{\beta}$, $H_{\gamma}$, $H_{\delta}$
fit the observed ones, because the $Y$ content influences the shape of the Balmer lines. 
Additionally, it is important to ensure that the theoretical profiles for the lines
$CaI~\lambda~4226$\AA\ and $CaII~\lambda\lambda~3933,~3968$\AA\ fit well the observed ones
for a given Ca content. If the ionization balance between $CaI$ and $CaII$ is reached,
our calculations are considered to be near completion.
  In Fig.~2 (a) we present isochrones for different ages and $Y$. 
The best-fit age and He abundance for NGC~2419 are
 $\log t=10.1$ and $Y=0.26$.
In Fig.~2 (b)  we show the theoretical line profiles 
$H_{\beta}$ and $H_{\delta}$  compared to the observed ones.
When reducing the age, the temperature of stars increases, and
 the Main sequence turn-off point (MSTO) gets bluer. 
This leads to an increased contribution of early-type Main Sequence $F$-dwarfs 
to the integrated light of the cluster. These stars have strong Balmer lines.
As a result of reducing the age,  all $HI$ lines appear to be more 
intense simultaneously in the doppler cores and in the wings.
On the other hand, increasing the He abundance does not influence strongly the 
shape of the isochrone (position of the Main Sequence, sub-giant and giant branches)
but it increases the luminosity of hot HB stars. 
It is known that the doppler cores of the Balmer lines are deep in the spectra 
of such stars, because gas pressure is not strong in their atmospheres. 
Therefore,  increasing $Y$ leads to a sharp increase of the central intensities
of the Balmer lines and a weakening of their wings. Note that 
 $H_{\delta}$ increases less than $H_{\beta}$ (see Fig.~2 (b)).
In summary,  $t$ and $Y$ influence differently the $HI$ line profiles, 
and this fact allows us to disentangle these two parameters.

Finally, we obtained $\log t=10.10\pm0.05$ and $Y=0.26\pm0.02$. With these values
the central depths  and the equivalent widths ($W_{\lambda}$) for the observed
and theoretical HI line profiles agree with an accuracy of $2\%$ and $5\%$,
respectively.  The obtained age of NGC~2419 agrees
well with the results of published photometric studies.
Bellazzini et al., 2012 analyzed narrow-band $HST$ stellar photometry.
They argue that the age of the cluster is in the range
$\log t=10.04$--$10.13$ using  isochrones of  Dotter et al. (2007) with
[Fe/H]$=-2.1$, $[\alpha/Fe]=0.4$, $\log t=10.08$ and of  
 Pietrinferni et al. (2004) with [Fe/H]$=-1.8$, $[\alpha/Fe]=0.0$, $\log t=10.11$.
The same age range $\log t=10.08$--$10.11$ was obtained earlier by 
Harris et al. (1997) using HST stellar photometry data.

\section{CHEMICAL COMPOSITION OF NGC~2419}

We determined the  mean chemical abundances in stellar atmospheres 
in a regular way by fitting the observed line profiles with theoretical ones.
However the moderate spectral resolution did not allow us to determine the abundances
using individual lines and molecular bands merging in blends.
This is why we determined the abundance of each element taking into account the 
agreement between the observed and theoretical spectra in all
blends containing its lines in the studied spectral range. 
In this fitting process all chemical elements were divided into 5 groups:

1) This group is composed of elements ($Fe$, $C$, $Ca$) with lines and molecular bands
 dominating in spectra. It is possible to derive unambiguous estimates of the abundances 
using these lines. They are accurate (up to  $0.1$ dex). Comparison of the observed 
and the theoretical spectra calculated for different  $[Fe/H]$ is shown in Fig.~3.

2) This group consists of elements ($N$, $Mg$, $Ti$, $Cr$, $Co$, $Mn$, $La$) 
with weak and blended lines and molecular bands. Their abundances were 
derived taking into account the  abundances for the elements of
the first group derived by us. The accuracy of abundances in this case is lower
 ($ \sim 0.2$ dex).

3) This group consists of elements with spectroscopic features too weak
 to be detected in the studied spectral range.
However, they influence the molecular and ionization equilibrium of other elements.
It is possible to obtain their approximate abundances from the analysis of this equilibrium.
For example, we argue that part of the C atoms forms the  $CO$ molecule if 
the O abundance is  enhanced. This reduces the intensity of the molecular bands
$CN$ and $CH$. Therefore, testing the equilibrium between different observed
and theoretical intensities of these bands allows one to estimate abundances of all $CNO$
group elements. The $CNO$ abundances are interdependent.

4) Sodium ($Na$) with  lines strongly distorted by interstellar components
 was considered in this group.

5) This group consists of elements ($Al$, $Si$, $V$, $Ni$) which  do not 
have lines detectable at medium resolution. 
However, they influence the equilibrium of electrons in stellar atmospheres. 
Abundances of these elements were considered to be equal
to that of other elements of this group.

The microturbulent velocity $\xi_{turb}$ was considered to be equal for all stars
of the cluster. We derived it from the best fit of weak and strong Fe lines
in the observed spectra.
Comparison between the observed and theoretical spectra calculated 
with different $\xi_{turb}$ and $[Fe/H]$ is shown in Fig.~4.
As a result of this analysis, a mean microturbulent velocity
was defined to be $\xi_{turb} = 2.1 \pm 0.2$ km/s.
Note that it is not quite correct to use such value $\xi_{turb}$,
bearing in mind a variety of stellar luminosities and temperatures in the cluster. 
Nevertheless, our calculations have shown that the hottest MSTO stars 
($T_{eff} \approx 6600$K, $\log~g \approx 4.0$) and red giant branch (RGB) stars
($T_{eff}=4100$--$4700$K, $\log~g=0.2$--$0.8$)
are the main contributors to the observed spectrum. For such stars our 
adopted value of $\xi_{turb}$ corresponds well to the typical  values found 
in the literature (Cayrel et al., 2004, Luck \& Heiter, 2005).

A final comparison of the observed and theoretical spectra is shown in Fig.~5. 
The obtained abundances are presented in Fig.~6
and Table~2 in comparison with the data from Cohen et al. (2011)
and Shetrone et al. (2001).

One may conclude that the derived chemical composition of NGC~2419 
agrees qualitatively with the data published for individual stars in the cluster.
 Cohen et al. (2011)  obtained on average $0.5$ dex higher O and Mg 
abundance for  red giants in NGC~2419. 
The abundances of other elements agree with our estimates within the errors ($0.2$ dex). 
It is worth noting that the O abundance was estimated indirectly  
by us, using the influence of this element on the molecular bands  $CH$, $CN$.
This estimate may not be very accurate.
Note that  Cohen et al. (2011) determined $[O/H]$ studying the infrared triplet 
$OI~\lambda\lambda~7771,~7772,~7774$\AA\ which is influenced strongly by non-LTE effects.
We suggest that to obtain an accurate O abundance it is important to resort to non-LTE analysis.
We derived the abundance of Mg using a strong not-blended line $Mg~\lambda~5183$\AA\
and a total flux distribution in the range of the index $MgH$ ($\sim 5050 \div 5300$\AA).
The abundance of Mg in Cohen et al. (2011) was determined at high spectral resolution 
using strong lines including $Mg~\lambda~5183$\AA.
We calculated non-LTE corrections for $Mg1$ in atmospheres of different stars of 
the cluster and obtained values less than $\Delta [Mg/H]=0.1$ dex. The reason for  
the high $[Mg/H]$ found in the literature was not clear to us. It is worth noting that there are
significant variations of the Mg abundance in the red giants studied by Cohen et al. (2011), 
Shetrone et al. (2001)
and differences in the methods of measuring and the theoretical calculation of strong $MgI$
lines in the literature and in our analysis. When we finished writing this paper, 
papers by Mucciarelli et al. (2012) and Cohen \& Kirby (2012) were published. 
A  large dispersion of $ Mg$ and $Ca$ abundances  for  RGB stars in NGC~2419 was discovered
by these authors and may partially explain the reason of the low  $[Mg/H]$ derived by us.

The chemical composition of NGC~2419 has features 
 distinguishing it from typical old GCs, or from a number of field stars in our Galaxy
(Borkova \& Marsakov, 2005, Pritzl \& Venn (2005), Venn et al., 2004, 
see also discussion in Cohen et al. (2011)).
According to the present theory of chemical evolution for our Galaxy,
(see the classical results of  Samland (1998) and later studies of chemical
evolution of Alibes et al., 2001 and Kobayashi et al., 2006)
there should be overabundances of  $[C/Fe]$, $[O/Fe]$, $[Mg/Fe]$, $[Si/Fe]$, $[Ca/Fe]$, 
and $[Ti/Fe]$ of an order $0.3$--$0.4$ dex for stars with metallicity [Fe/H]$=-2.2$ dex.
Additionally, a deficit of [Na/Fe], [Al/Fe] of the order $0.2$ dex is also expected.
Other elements (including $Zn$) have solar X/Fe abundance ratios .  
Table~2 shows that both estimates of the C abundance for NGC~2419  
indicate its extreme deficit  $[C/Fe] < 0.7$ dex which cannot be explained by 	
supply  of the material  synthesized in the last stages of evolution on its surface.
Overabundances of $[Ca/Fe]$, $[Ti/Fe]$ in stars of the cluster ($0.1$--$0.2$ dex) 
appear to be much less than the ones suggested by Samland (1998). 
$[Mg/Fe]$ has a value close to the Solar one.

\section{CONCLUSION}

We have carried out theoretical population synthesis modeling for NGC~2419, 
an unusual GC in the outer Galactic halo, using medium-resolution spectra. 
To solve the task
we developed a new method which may be applied to medium-resolution spectra 
($R \ge 2500$) with high signal-to-noise ratio ($S/N \ge 100$) in a wide spectral range
$ > 1500$ \AA. The efficiency of this method is higher for higher spectral resolution,
wider spectral range and larger $S/N$.
The example considered in this paper is almost limiting for these parameters.
We used {\it as an initial approximation} the isochrone, the IMF, and
 distribution the cluster stars according to their masses, radii, and $ log~g $
using the published theoretical isochrone corresponding best to the observed $CMD$ 
(Dalessandro et al., 2008; Bellazzini et al., 2012).
 We summed the  synthetic blanketed stellar spectra calculated 
using the models of stellar atmospheres according to the Chabrier (2005) IMF
which gives a good approximation of the stellar LF  found in the literature for 
NGC~2419 (Bellazzini et al., 2012). Fitting of the observed spectra  by the theoretical 
ones was made iteratively in several stages. 
Finally, we adopted the chemical composition of stars corresponding to the selected 
isochrone. To determine the age and helium content of the stars we analyzed mainly 
the depth and the shape of the wings of HI lines, because these are 
independent of other parameters. Then we varied the Fe abundance, since these lines
dominate the optical spectrum, even at low metallicity.
After optimization of the aforementioned main parameters (age, $ log~g $, He and Fe abundance),
we fitted the C, Mg, and Ca abundances. We found that  molecular bands $ MgH$,
 $ CH$ and $ CaII$ lines are some of the most prominent features in the spectrum of
NGC~2419. To derive abundances of other chemical elements it was necessary
to analyze blends of many lines. Finally, we managed to estimate abundances of 14
elements including 3 alpha-elements (Ca, O, Mg) and Ti. We obtained 
age=12.6 Gyr, [Fe/H]$=-2.25$ dex, $Y=0.26$ and [$\alpha$/Fe]$= 0.13$ as the mean of
$[O/Fe]$, $[Mg/Fe]$ and $[Ca/Fe]$. The derived chemical composition (except for the
alpha-elements) agrees well with the published one, which was obtained using high-resolution 
spectra for 8 RGB stars of the cluster (Cohen et al. (2011), Shetrone et al. (2001)), 
and confirms the conclusion about the chemical peculiarity of NGC~2419. 
The abundance of Mg ($[Mg/Fe]=-0.05 \pm 0.15$) derived by us agrees within the errors
 with a mean value for 49 RGB stars obtained by Mucciarelli et al. (2012) 
($[Mg/Fe]=0.05 \pm 0.08$) using high-resolution spectroscopy.

The method presented here employs information from all the stars in the cluster, 
and may be successfully applied to integrated-light spectroscopic studies of extragalactic GCs.
It is still impossible to obtain deep $CMDs$, or high-resolution, high $S/N$ spectra
for such objects. In these cases, an initial approximation for the age may be derived
 using Lick indices, or full spectral fitting using simple stellar population models
(see e.g. Sharina et al. (2010) and references therein).

ACKNOWLEDGEMENTS.
{\scriptsize
S.M.E. acknowledges partial support by the Russian Foundation for Basic Research (RFBR) 
grant 11-02-00639, and a Russian-Ukrainian RFBR grant 11-02-90449; 
 support of the Ministry of Education and Science of the Russian
Federation, the contract 14.740.11.0901 and proposal 2012-1.5-12-000-1011-004.
 S.V.V. acknowledges support by the RFBR grant 10-02-01145-a. We thank Th. H. Puzia for interesting comments.
}

\newpage
\begin{table}[htbp]
\begin{center}
\caption{Journal of spectroscopic observations.}
\begin{tabular}{|l|c|c|c|}
\hline
Object    & Data       & Exposition, sec.  & Slit position, degr. \\
\hline
NGC~2419    & 1.12.08    & 4x600, 300        & 111 + scan             \\
           &            & 2x1200            & 180                    \\
           & 3.12.08    & 2x1800, 600       & 148                    \\
HR1544     & 1.12.08    & 2                 & 180                    \\
HR8924     & 1.12.08    & 1                 & 180                   \\
HR1805     & 1.12.08    & 2                 & 180                   \\
HR2002     & 1.12.08    & 2                 & 180                    \\
HR3422     & 1.12.08    & 2                 & 180                    \\
HR3418     & 3.12.08    & 2                 & 180                    \\
HR3905     & 3.12.08    & 2                 & 180                    \\
\hline
\end{tabular}
\end{center}
\end{table}

\begin{table}[htbp]
\begin{center}
\caption{Mean abundances $[X/H]$, in dex and corresponding dispersions $\sigma$, in dex 
for different chemical elements in NGC~2419 according to our work (columns 2 and 3),
and those from  Cohen et al. (2011) for red giant members of the cluster (columns 4 and 5)
and Shetrone et al. (2001) (column 6)}\label{tab1}
\begin{tabular}{|l|c|c|c|c|c|}
\hline
elem. & $[X/H]$ & $\sigma$ & $[X/H]$ & $\sigma$ & $[X/H]$ \\
\hline
He    & -0.05 & 0.05 &       &      & \\
C     & -2.65 & 0.10 & -2.82 & 0.25 & \\
N     & -2.00 & 0.20 &       &      & \\
O     & -1.90 & 0.30 & -1.38 & 0.10 & \\
Na    & -2.10 & 0.15 & -2.03 & 0.31 & \\
Mg    & -2.30 & 0.15 & -1.82 & 0.36 & -2.02 \\
Si    &       &      & -1.69 & 0.12 & \\
K     &       &      & -1.54 & 0.28 & \\
Ca    & -2.15 & 0.10 & -1.97 & 0.07 & -2.21 \\
Sc    &       &      & -1.95 & 0.14 & \\
Ti    & -2.05 & 0.20 & -1.82 & 0.16 & -2.10 \\
V     &       &      & -2.09 & 0.09 & \\
Cr    & -2.15 & 0.15 & -2.35 & 0.09 & \\
Mn    & -2.15 & 0.15 & -2.40 & 0.07 & \\
Fe    & -2.25 & 0.10 & -2.12 & 0.07 & -2.32 \\
Co    & -2.25 & 0.15 & -2.03 & 0.08 & \\
Ni    &       &      & -2.17 & 0.03 & \\
Cu    &       &      & -2.13 & 0.01 & \\
Sr    &       &      & -2.35 & 0.13 & \\
Y     &       &      & -2.48 & 0.09 & -2.29 \\
Ba    & -2.10 & 0.20 & -2.23 & 0.14 & -2.47 \\
La    & -2.25 & 0.20 & -2.07 & 0.30 & \\
Nd    &       &      & -2.12 & 0.23 & \\
Eu    &       &      & -1.82 & 0.15 & \\
\hline
\end{tabular}
\end{center}
\end{table}

\newpage
\begin{figure*}[htbp]
\vspace*{220mm}
\includegraphics{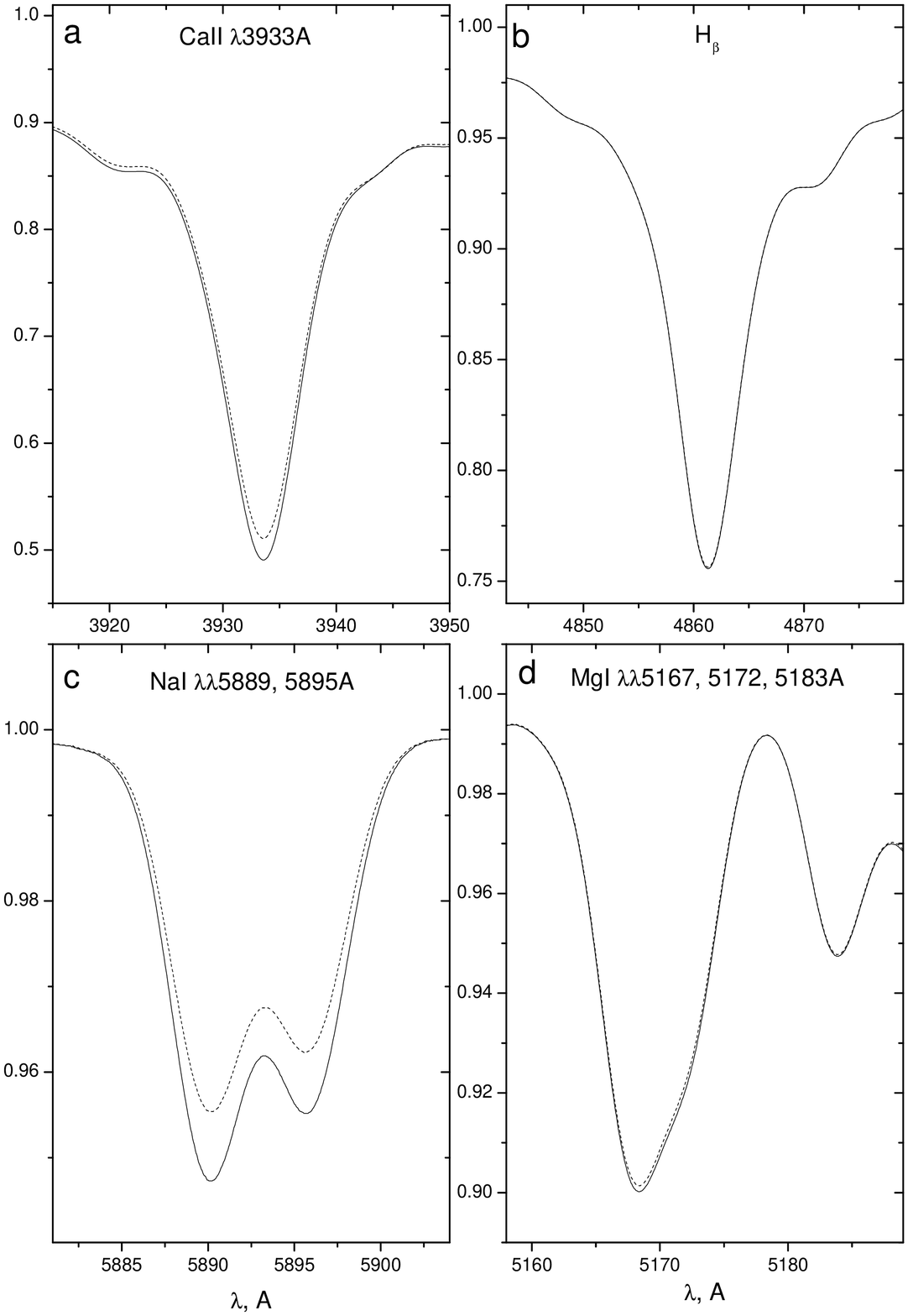}
\caption{Influence of non-LTE effects on the lines $CaII$ (a), $HI$ (b), $NaI$
(c) and $MgI$ (d) in the integrated spectrum of NGC~2419. The profiles
were calculated for the  parameters and 
chemical composition of NGC~2419 derived by us (see Sec.~3)
using the LTE approach (dashed lines) and
taking into account non-LTE effects (thick lines).}
\label{fig1}

\end{figure*}

\newpage
\begin{figure*}[htbp]
\vspace*{220mm}
\includegraphics{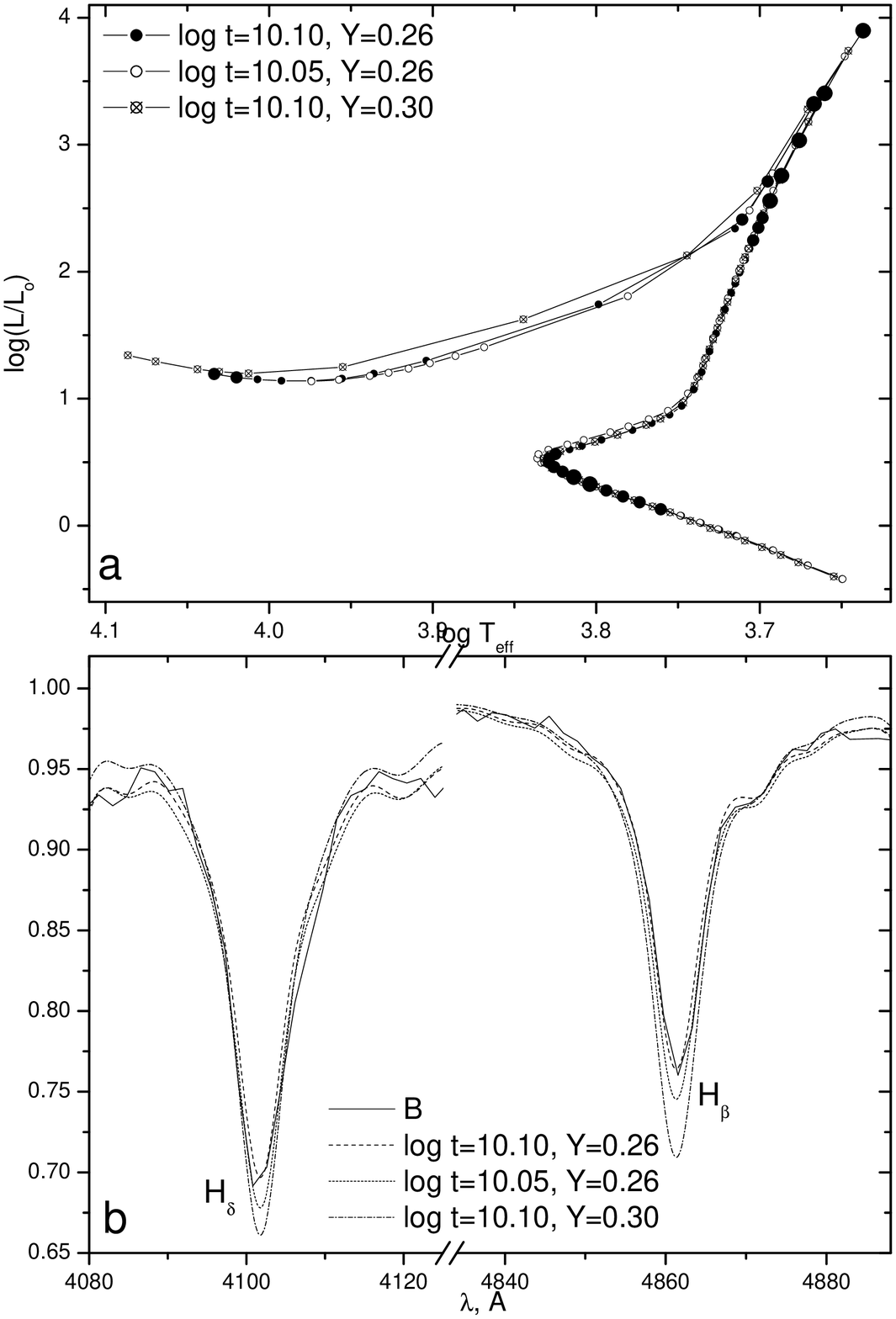}
\caption{(a) Theoretical isochrones from Bertelli et al. (2009) in the diagram $ log(T_{eff})$ - $ log~g$ 
and (b) $ H_{\beta}$, $ H_{\delta}$ line profiles in the integrated spectrum of NGC~2419
 for a set of the cluster parameters $t$ and $Y$ for $Z=0.0001$. 
Places of the isochrone $\log t=10.10$, $Y=0.26$ with different contributions to the total flux 
at $ \lambda = 5000$\AA\ are shown by circles of different size: less than $1\%$,
$1\% \div 3\%$ and $ > 3\%$.}
\label{fig2}
\end{figure*}

\newpage
\begin{figure*}[htbp]
\vspace*{130mm}
\includegraphics{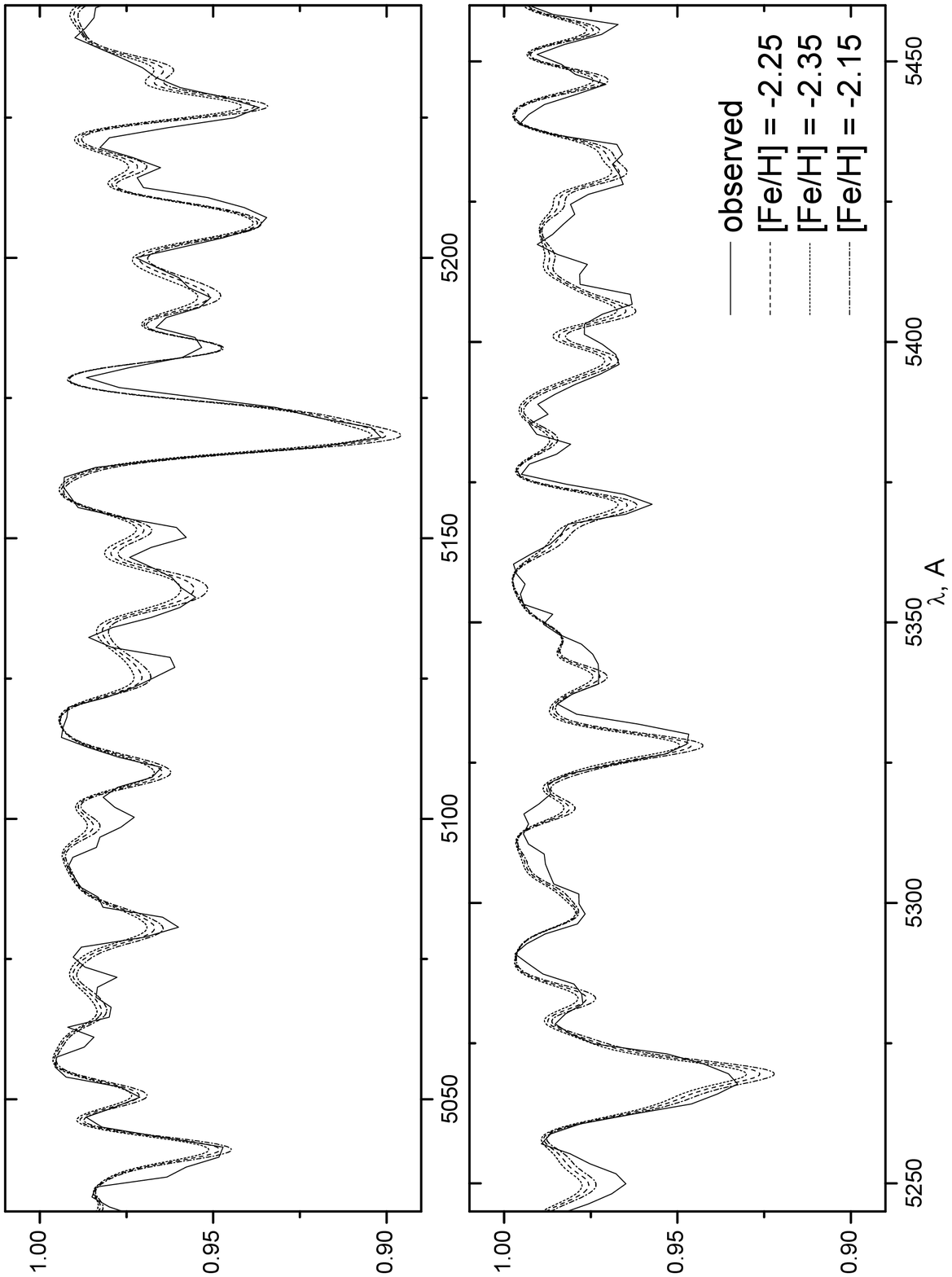}
\caption{Determination of the Iron abundance $[Fe/H]$. The observed (thick lines) and 
theoretical (dashed, dashed-dot  and long-dashed lines) spectra of NGC~2419 
calculated for different $[Fe/H]$ without changing the abundance of other elements
(See Sec.~4).}
\label{fig3}
\end{figure*}
\newpage
\begin{figure*}[htbp]
\vspace*{130mm}
\includegraphics{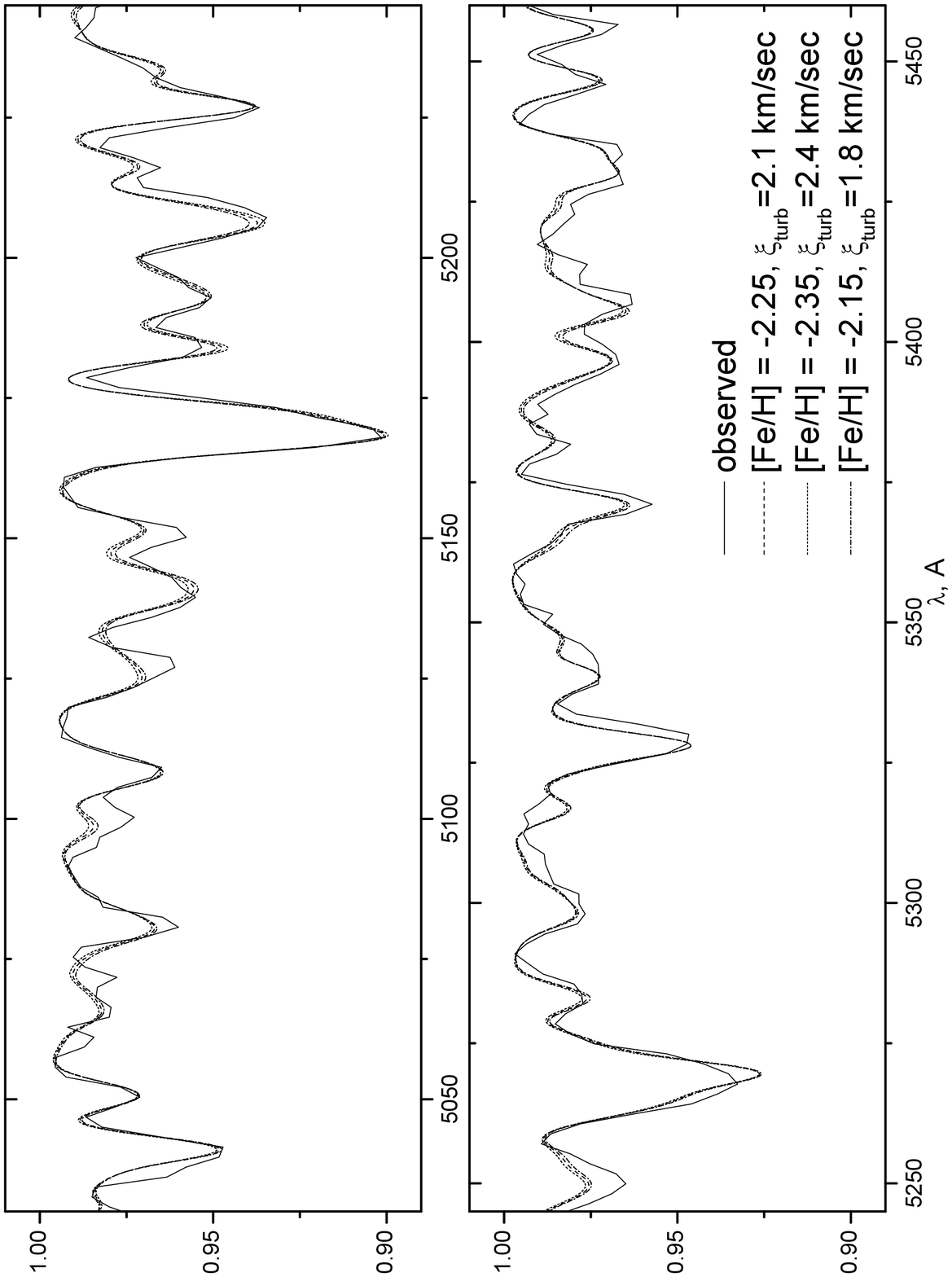}
\caption{Determination of the microturbulent velocity  $\xi_{turb}$. 
The observed (thick lines) and theoretical (dashed, dashed-dot  and long-dashed lines)
spectra of NGC~2419  calculated for different $\xi_{turb}$ and $[Fe/H]$.}
\end{figure*}

\newpage
\begin{figure*}[htbp]
\vspace*{220mm}
\includegraphics{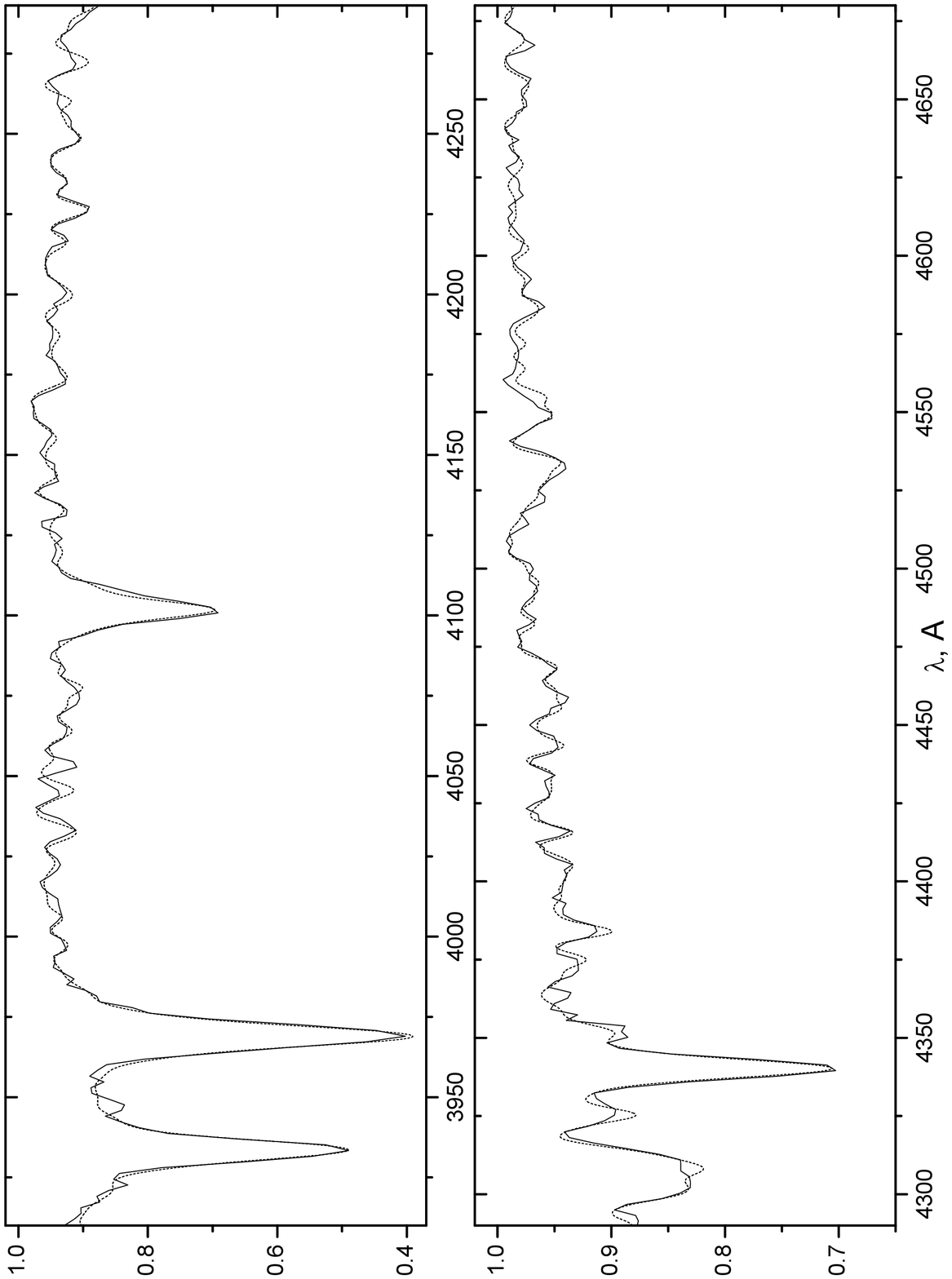}
\includegraphics{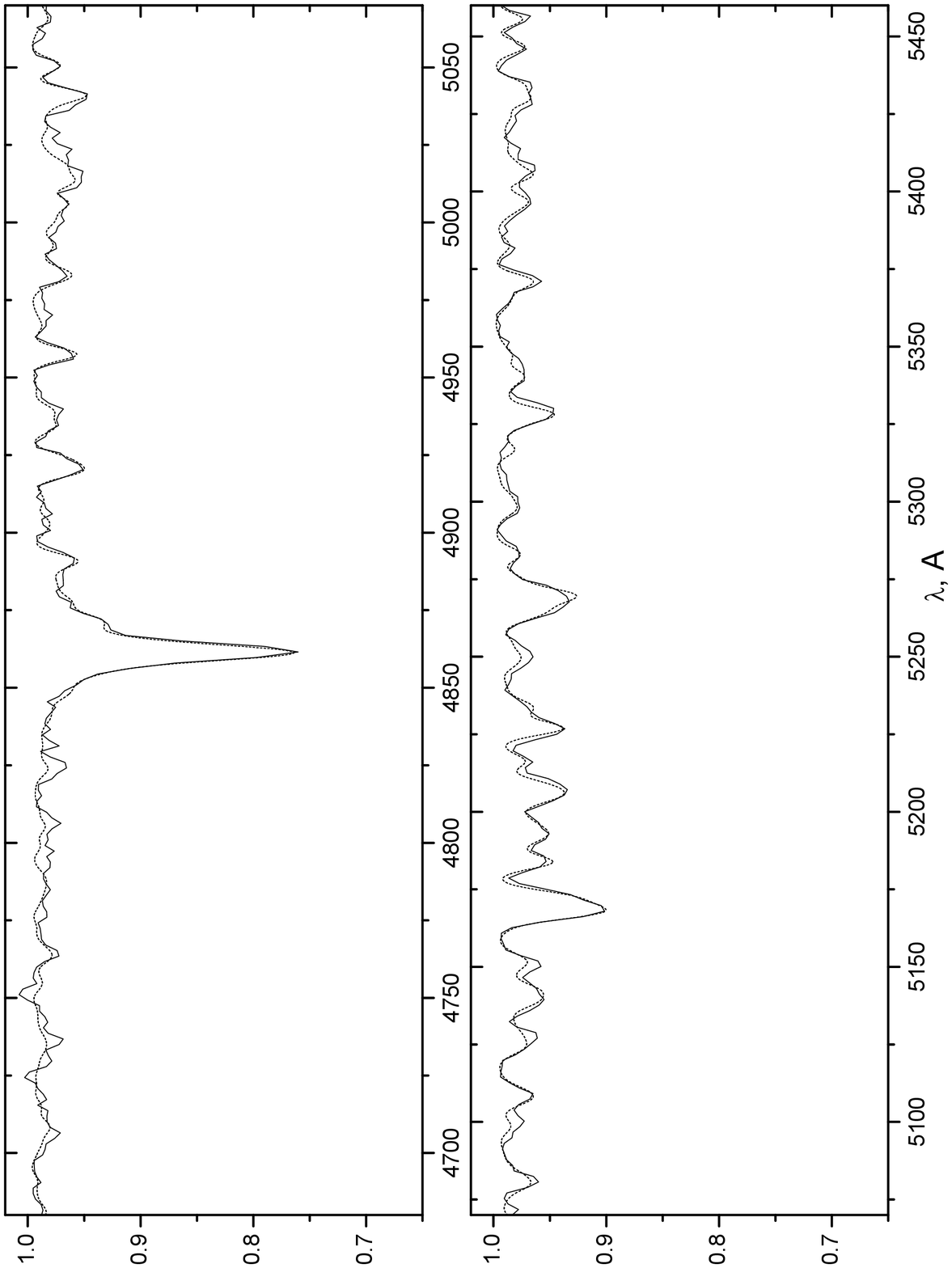}
\caption{The observed (thick lines) and theoretical (dashed lines) spectra of NGC~2419
in the range $\lambda = 3910$--$4680 $\AA\ (top) and  $\lambda = 4680$--$5460 $\AA\ (bottom). 
The theoretical spectrum was calculated  
according to the obtained parameters and chemical composition (see Sec.~3.4).}
\end{figure*}

\newpage
\begin{figure*}[htbp]
\vspace*{120mm}
\includegraphics{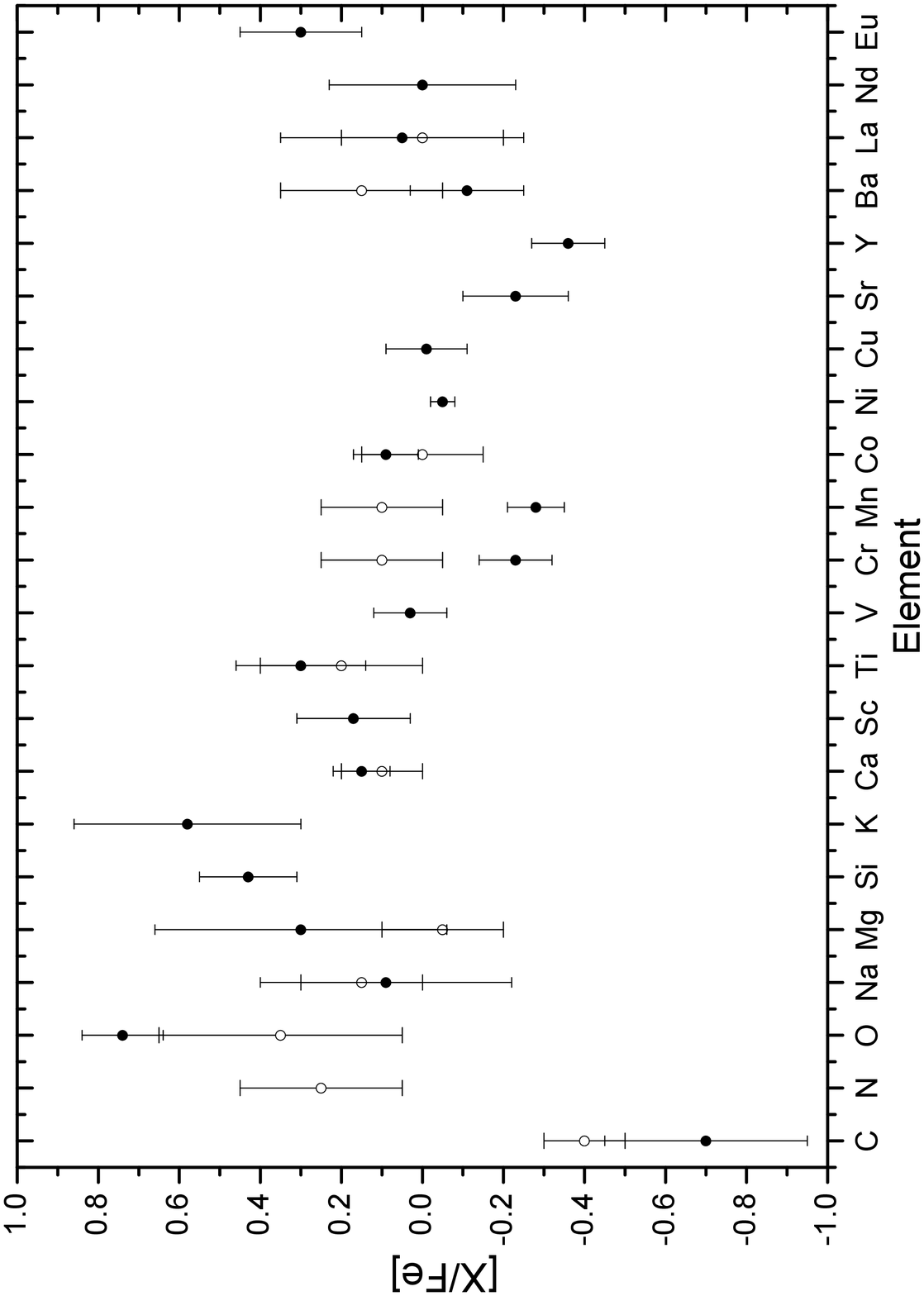}
\caption{Chemical abundance of NGC~2419 derived in our paper (open circles) in comparison with
the results of  Cohen et al. (2011).}
\end{figure*}

\end{document}